\documentclass[letterpaper]{article} 
\usepackage{aaai24}  
\usepackage{times}  
\usepackage{helvet}  
\usepackage{courier}  
\usepackage[hyphens]{url}  
\usepackage{graphicx} 
\usepackage{natbib}  
\usepackage{caption} 
\usepackage{algorithm}
\usepackage{algorithmic}
\usepackage{multirow}

\usepackage{amsmath} 
\usepackage{diagbox} 

\usepackage{newfloat}
\usepackage{listings}
\DeclareCaptionStyle{ruled}{labelfont=normalfont,labelsep=colon,strut=off} 
\floatstyle{ruled}
\newfloat{listing}{tb}{lst}{}
\floatname{listing}{Listing}
\title{Enhancing Zero-Shot Multi-Speaker TTS with Negated Speaker Representations}
\author{
    Yejin Jeon\textsuperscript{\rm 1},
    Yunsu Kim\textsuperscript{\rm 2},
    Gary Geunbae Lee\textsuperscript{\rm 1,\rm 3}\\
}
\affiliations{
    \textsuperscript{\rm 1}Graduate School of Artificial Intelligence, POSTECH, Republic of Korea\\
    \textsuperscript{\rm 2}aiXplain Inc. Los Gatos, CA, USA\\
    \textsuperscript{\rm 3}Department of Computer Science and Engineering, POSTECH, Republic of Korea\\

    \{jeonyj0612, gblee\}@postech.ac.kr, yunsu.kim@aixplain.com
}

\usepackage{bibentry}

\begin{document}

\maketitle

\begin{abstract}
Zero-shot multi-speaker TTS aims to synthesize speech with the voice of a chosen target speaker without any fine-tuning. Prevailing methods, however, encounter limitations at adapting to new speakers of out-of-domain settings, primarily due to inadequate speaker disentanglement and content leakage. To overcome these constraints, we propose an innovative negation feature learning paradigm that models decoupled speaker attributes as deviations from the complete audio representation by utilizing the subtraction operation. By eliminating superfluous content information from the speaker representation, our negation scheme not only mitigates content leakage, thereby enhancing synthesis robustness, but also improves speaker fidelity. In addition, to facilitate the learning of diverse speaker attributes, we leverage multi-stream Transformers, which retain multiple hypotheses and instigate a training paradigm akin to ensemble learning. To unify these hypotheses and realize the final speaker representation, we employ attention pooling. Finally, in light of the imperative to generate target text utterances in the desired voice, we adopt adaptive layer normalizations to effectively fuse the previously generated speaker representation with the target text representations, as opposed to mere concatenation of the text and audio modalities. Extensive experiments and validations substantiate the efficacy of our proposed approach in preserving and harnessing speaker-specific attributes vis-\`{a}-vis alternative baseline models.
\end{abstract}

\section{Introduction}
Recent neural network-based models have revolutionized the field of text-to-speech (TTS), generating natural speech that closely resembles human-like qualities \cite{Taco2, FastSpeech2}. This has fueled a demand for \textit{personalized} TTS capabilities, particularly voice imitation using multi-speaker TTS. Yet, the effective implementation of such a task presents formidable challenges. First, multi-speaker TTS models must be able to adapt to diverse speakers, including those unseen in the data, while maintaining fidelity across all text inputs. Second, there exist practical limitations in gathering sufficient speech data for each speaker for accurate imitation by TTS models. Thus, the ability to accurately replicate the voices of both seen and unseen speakers with minimal audio data samples, or even from just one audio input, has gained paramount significance.

One avenue of investigation involves fine-tuning a pretrained model initially trained on an extensive multi-speaker dataset, with a limited set of speech samples of the desired target speaker \cite{FineTune1, FineTune2}. While this few-shot approach has demonstrated good speaker adaptability, it is not impervious to inherent limitations. In particular, it necessitates additional training tailored to each target speaker to ensure accurate synthesis of speech in their specific voice. Moreover, when confronted with a scarcity of audio samples for fine-tuning, the resulting voice quality exhibits a noticeable degradation in speaker similarity.

In response to the challenges observed in few-shot synthesis, another branch of research have focused on extracting target speaker embeddings from a single reference audio sample in a zero-shot manner. Some investigations utilize x-vectors that are generated from pretrained speaker verification models, which in turn are trained with large voice datasets \cite{Xvector1}. However, not only does this method entail substantial data requisites, but also experiences limited speaker fidelity for out-of-domain speakers. Consequently, other endeavors have gravitated towards strategies of reduced data expenditure while conceptualizing speaker attributes as latent characteristics. Specifically, target speaker attributes are learned through unsupervised methods like Global Style Tokens (GST) \cite{GST-2} or variational autoencoders (VAE) \cite{VAE-2}. While a certain degree of speaker adaptation is achievable through the aforementioned methods, speaker-dependent factors are not sufficiently disentangled from other components such as the uttered linguistic content. Thus, models learn to extract speaker embeddings that include undesired content information. Not only does this lead to discrepancies in speaker timbre, but also towards unstable synthesis since the TTS model learns to rely on the contextual cues from the intended target input text, as well as those provided via the reference audio.

Towards accurate and robust multi-speaker TTS, we propose a novel zero-shot and end-to-end model that incorporates the following fundamental aspects. (1) We introduce for the first time, a negated feature learning strategy that effectively generates speaker representations by computing it as the differentials between the full representation of the reference audio and its linguistic contents. Not only does this strategy facilitate better disentanglement of target timbre information from other irrelevant speech factors, but also decrease content leakage that is introduced into target speaker representations. (2) We further implement multi-stream Transformers in order to generate multiple hypotheses such that various speaker attributes can be learned and preserved. By using attention pooling, a more optimal speaker representation is computed from the unification of various speaker hypotheses. (3) Knowledge of the ultimate speaker representations are injected into the backbone TTS pipeline using adaptive layer normalizations to enhance target text and speaker fusion. Through in-depth experimentation and verification, we demonstrate that our zero-shot unsupervised approach is more robust and achieves higher speaker similarity compared to previous baseline studies.

\section{Related Work}
\subsection{Controllable Speech Generation Tasks}
The pursuit of controllable speech generation has given rise to a range of diverse tasks. Among these, a pertinent task to multi-speaker TTS is voice conversion, wherein an utterance articulated by a source speaker is modified to make it sound as though spoken by a different target speaker. Since this transformation requires that the linguistic contents of the source audio remain unchanged, this necessitates the availability of parallel speech data where diverse speakers articulate identical text. Given this setting, several investigations have concentrated on achieving alignments between parallel spectral features by employing techniques like dynamic frequency warping \cite{Related1}. Nevertheless, the scarcity of readily available parallel datasets has prompted certain studies to resort to non-parallel data. Yet, these endeavors are constrained to voice conversion between a predefined set of speakers \cite{Related2, Related3} and are contingent on one-hot speaker representations \cite{Related4, Related5}. In contrast, multi-speaker TTS demands precise speaker modelling that can generalize to any textual inputs, including those outside the established linguistic domain of the source audio. For this purpose, unlike unimodal acoustic modeling seen in voice conversion, multi-speaker TTS requires dual-modality modeling, and fusion of text and audio domains. Furthermore, model training is not confined to parallel speaker datasets.

\subsection{Attribute-Specific Disentanglement}
Due to its potential to govern the generative process of outcomes, the disentanglement of target features is a pivotal aspect that transcends various research domains. Foremost in this pursuit were endeavors within computer vision for style transfer; the semantic contents of the source image are imbued with the stylistic intricacies of the target image using pairwise correlation matrices \cite{Related6}. This seminal work initiated an array of investigations such as those using different variants of instance normalizations \cite{CondInstNorm-Related, Ulyanov-Related}. This thematic trend subsequently diffused into adjacent areas such as machine translation, where each source-target translation pairs were annotated by the labels of the translators who generated that particular translation \cite{Related7}. By doing so, translations could be generated in a translator-specific linguistic style pertaining to their grammatical preferences and verbosity. Similarly, the domain of speech has witnessed a large amount of work investigating the manipulation of both discrete and continuous attributes such as speaker accents \cite{Accent-Related}, emotion nuances \cite{Emotion-Related}, and prosodic variations \cite{Intonation-Related}, each forming their own research sub-domains. 

Towards multi-speaker TTS, speaker modeling from a reference audio is mainly conducted in two ways: 1) direct, fixed-length extraction of timbral attributes via a speaker encoder; and 2) additional disentanglement of target speaker embeddings from other observable factors. For the former, many works assume that timbre is a global feature in speech, and thus utilize coarse-grained methodologies such as pretrained wav2vec models \cite{wav2vec-Related}, hierarchical GSTs \cite{HierGST}, and multivariate Gaussian mixture VAEs \cite{GMVAE-Related}. The latter paradigm involves maximizing the distance between the directly generated speaker vectors with that of contextual representations. This is done by first pretraining the backbone TTS system with a single speaker dataset, which allows for the content encoder to encapsulate just contextual information. A Mutual-Information Neural Estimator (MINE) is then used to maximize the lower bound for mutual information through Donsker-Varadhan representation \cite{Mutual1-Related}. Similarly in \cite{Renyi-Related}, R\'enyi Divergence is used to create more nuanced speaker representations by establishing distance between timbre and content information. 

\begin{figure}[t]
    \centering
    \includegraphics[width=0.94\linewidth, height=11.3cm]{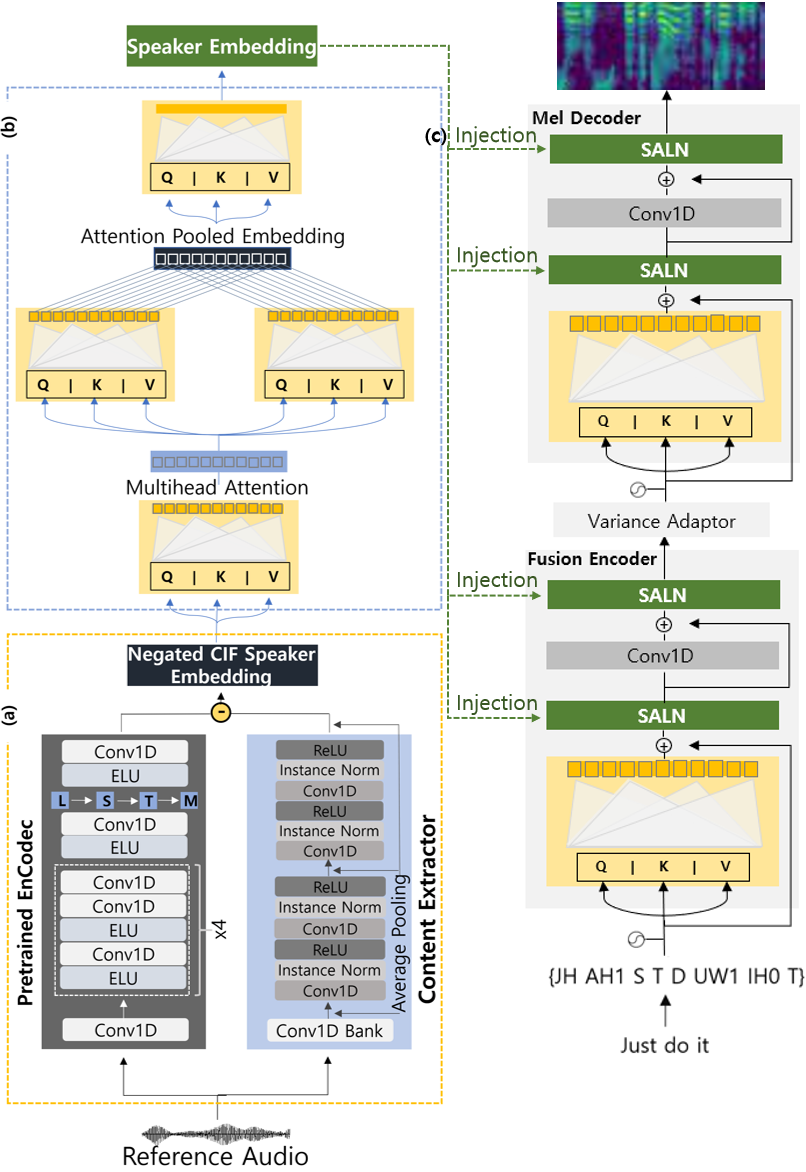}
    \caption{Main architecture: (a) Negated feature generation, which results in CIF embeddings, are passed to (b) multi-stream Transformers to formulate multi-perspective speaker representations, which are then (c) injected into the TTS backbone. Target speaker embeddings are fused together with input text representations at the fusion encoder, and re-injected into the mel decoder for further speaker preservation.}
    \label{fig:main_arch}
\end{figure}

\section{Methodology}
\subsection{Preliminaries}
Given training dataset $D = {(X_n, Y_n)}_{n=1}^{N}$ of $N$ samples and where $X_n$ and $Y_n$ form respective pairs of text and audio, the objective of conventional single-speaker TTS frameworks lies in generating acoustic representation $\hat{Y}_n$ that faithfully corresponds to the input text $X_n$. This process comprises of an acoustic model and a vocoder. The acoustic model is further consisted of three components: a linguistic encoder responsible for generating a contextual representation for $X_n$, an alignment module, and a generator that produces intermediate acoustic representations in the form of mel-spectrograms $\hat{M}_n$. The vocoder then converts $\hat{M}_n$ into $\hat{Y}_n$. In this work, we focus solely on the former acoustic model. 

For the baseline acoustic model, we leverage the non-autoregressive FastSpeech model \cite{FastSpeech2}. Specifically, a text sequence $X_n = \{x_1, x_2, ..., x_l\}$ of length $L$ is converted to a context embedding via linguistic encoder $\mathcal{E}_{text}$, which is a feed-forward Transformer block. The alignment module is formulated as the variance adapter, which adds variance information to help facilitate one-to-many mapping necessary for text-to-speech. The variance adapter, which is comprised of pitch, energy, and duration predictors, are each made up of a 2-layer, 1-D convolutional network followed by ReLU activation, layer normalizations, dropout, and a linear layer. The resulting embedding is then utilized by generator $\mathcal{E}_{mel}$, which has the same structure as $\mathcal{E}_{text}$, to produce $\hat{M}_n$ corresponding to $X_n$. 

Given the imperative to generate audio in the distinctive timbre of the intended speaker for multi-speaker TTS, the singular textual requisite for single-speaker TTS shifts to dual inputs (i.e., reference audio and text utterance to control speaker voice and the uttered content, respectively). Thus, in conjunction with the original text-to-audio training flow, we propose and incorporate a novel concurrent speaker-specific pipeline that estimates disentangled speaker-specific attributes from reference audio $Y_n$ in a zero-shot manner. This is explained in the subsequent sections.


\subsection{Negated Feature Extraction}
Unlike prior methodologies that rely on a single encoder block to directly learn speaker features \cite{Grad_StyleSpeech} and maximize the distance between speaker and content features \cite{Mutual1-Related, Renyi-Related}, we introduce a novel negation strategy that entails \textit{subtracting} content information from a comprehensive audio representation, which results in disentangled content-information-free (CIF) speaker-specific embeddings (Fig.~\ref{fig:main_arch}-(a)). While this arithmetic property has been validated in computer vision research for facial expression learning \cite{Facial-subtract}, this is the first time in which negation has been studied for zero-shot multi-speaker TTS. Thus, our negation strategy requires two distinct components: full representation of the reference audio, and the content embeddings extracted from the same acoustic input. In order to generate the total audio representation, we use a pretrained audio compression EnCodec model \cite{EnCodec}, as it is capable of compressing audio without loss of information into a compact representation\footnote{We used the official 24kHz EnCodec model from \url{https://github.com/facebookresearch/encodec}}. The reference audio undergoes a series of transformations, including passing through a 1D convolutional layer with a kernel size of 7, and then four convolution blocks. Each convolution block consists of non-linear ELU activations and a strided convolution with a kernel size twice that of the stride. Strides of sizes 2, 4, 5, and 8 are employed. 

Simultaneously, to extract content information, we employ instance normalization (IN) - a well-proven method in stylistic transfer in images \cite{CV_IN}, which normalizes artistic styles (e.g., textures, brushstrokes, etc.) so that only the underlying image content is preserved. Similarly, since timbre can be thought of as the \textit{style} applied to the text content, we implement IN in our content extractor to extract content information from the reference audio. This is done by independently calculating the mean $\mu_c$ and standard deviation $\sigma_c$ of each $W$-dim channel $c$ of the feature map $F$, which is generated from the previous convolution layer for each reference sample. $\mu_c$ and $\sigma_c$ are then used to normalize each element of $F_{cw}$. To maintain numerical stability, a small value $\epsilon$ is added as seen in Equation 2. IN can be mathematically formulated as follows:

\begin{equation} 
    \mu_c(M) = \frac{1}{W} \sum_{w=1}^{W} F_{cw}
\end{equation}

\begin{equation}
    \sigma_c(F) = \sqrt{\frac{1}{W} \sum_{w=1}^{W} (F_{cw} - \mu_c(F))^2 + \epsilon}
\end{equation}

\begin{equation}
    IN(F) = \frac{F_{cw} - \mu_c(F)}{\sigma_c(F)}
\end{equation}

Overall, the content extractor incorporates a convolutional bank, whose output is then passed onto two blocks of twice repeating series of convolution 1D, IN, and ReLU. Additional information in the form of average pooling \cite{VC_IN} is introduced between the two blocks.

\begin{figure*}[t]
    \centering
    \includegraphics[width=.29\textwidth, height=4.1cm]{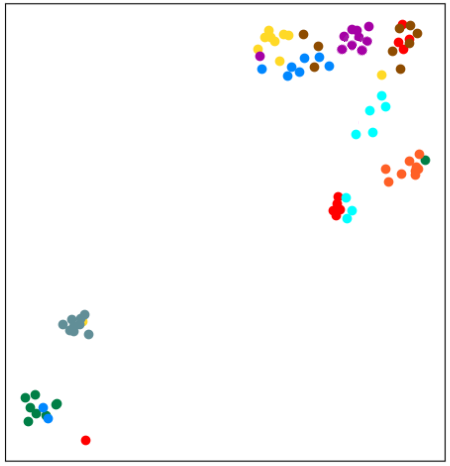}
    \includegraphics[width=.29\textwidth, height=4.1cm]{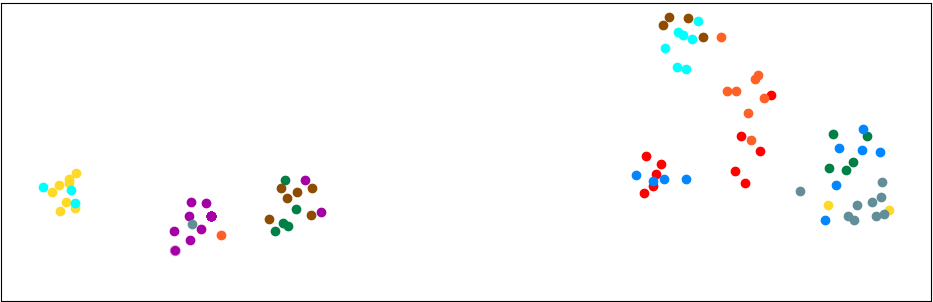}
    \includegraphics[width=.29\textwidth, height=4.1cm]{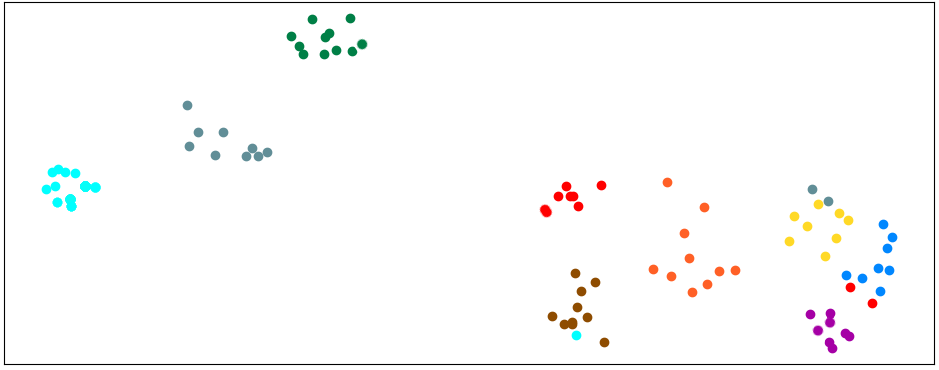}
  \caption{Comparative visualizations of distinct speaker embeddings. From left to right are baselines \citeauthor{Mellotron}, \citeauthor{SALN}, and our proposed model.}
  \label{fig:visualizations}
\end{figure*}

\subsection{Multi-stream Transformers}
In order to achieve accurate speech synthesis in any speaker's voice, it is crucial to capture various timbre patterns inherent in the input acoustic speaker prompt. However, existing approaches in multi-speaker TTS solely rely on a single linear information pathway to generate just one speaker embedding, which restricts the model from leveraging alternative hypotheses that capture different perspectives of the target speaker's timbre. Drawing inspiration from ensemble models \cite{Ensemble}, which combine multiple hypotheses together to enhance predictive performance, we utilize a multi-stream Transformer \cite{MultiStreamTrans} architecture originally used for machine translation (Fig.~\ref{fig:main_arch}-(b)).

\begin{table}[t]
\centering
\scalebox{0.91}{
\begin{tabular}{l|l|ccc}
\hline
\multicolumn{1}{l|}{\textbf{Model}} & \textbf{Speaker} &\textbf{MCD ($\downarrow$)} & \textbf{MOS ($\uparrow$)} & \textbf{WER ($\downarrow$)} \\\hline
\multirow{2}{*}{\citeauthor{Mellotron}}        
& Seen & 6.60 & 3.34 ± 0.20 & 21.82\% \\
& Unseen & 8.97 & 3.25 ± 0.19 & 26.76\% \\ \hline
\multirow{2}{*}{\citeauthor{SALN}}     
& Seen & 5.97 & 3.56 ± 0.19 & 17.21\% \\
& Unseen &9.52 & 3.51 ± 0.19 & 18.34\%  \\ \hline
\multirow{2}{*}{\textbf{Ours}} & Seen & 5.59 & 3.86 ± 0.18 & 15.37\% \\
& Unseen & 8.57 & 3.84 ± 0.18 & 16.14\% \\\hline
\end{tabular}
}
\caption{Main results for baselines and proposed models. Better performance is determined through lower MCD and WER, and higher MOS scores.}
\label{tab:main_results}
\end{table}

In the previous subsection, a novel negation strategy generated a CIF speaker representation. This representation is subsequently fed into a standard Transformer block to effectively capture long-range dependencies, and then passed to the multi-stream Transformer. Unlike the conventional Transformer \cite{OrgTransformer}, our design features two separate and parallel encoder streams. Each stream conducts independent computations on the full CIF embedding, enabling the exploration and preservation of distinct timbre hypotheses. Furthermore, a Transformer stream employs identical blocks as those found in the original Transformer encoder. All blocks feature 2 heads for the multi-head attention and is of depth 1, ensuring that a single Transformer encoder block is used for each parallel stream. The resulting speaker embedding has a dimension of 128. While the construction of these multiple alternative perspectives is essential, leveraging these hypotheses effectively is another important aspect for improving timbre predictions. Towards this end, we conduct attention pooling across all tokens occupying the same position within each independent Transformer stream, facilitating integration of information from multiple perspectives.

\begin{figure}[t]
    \centering
    \includegraphics[width=0.88\linewidth, height=5.4cm]{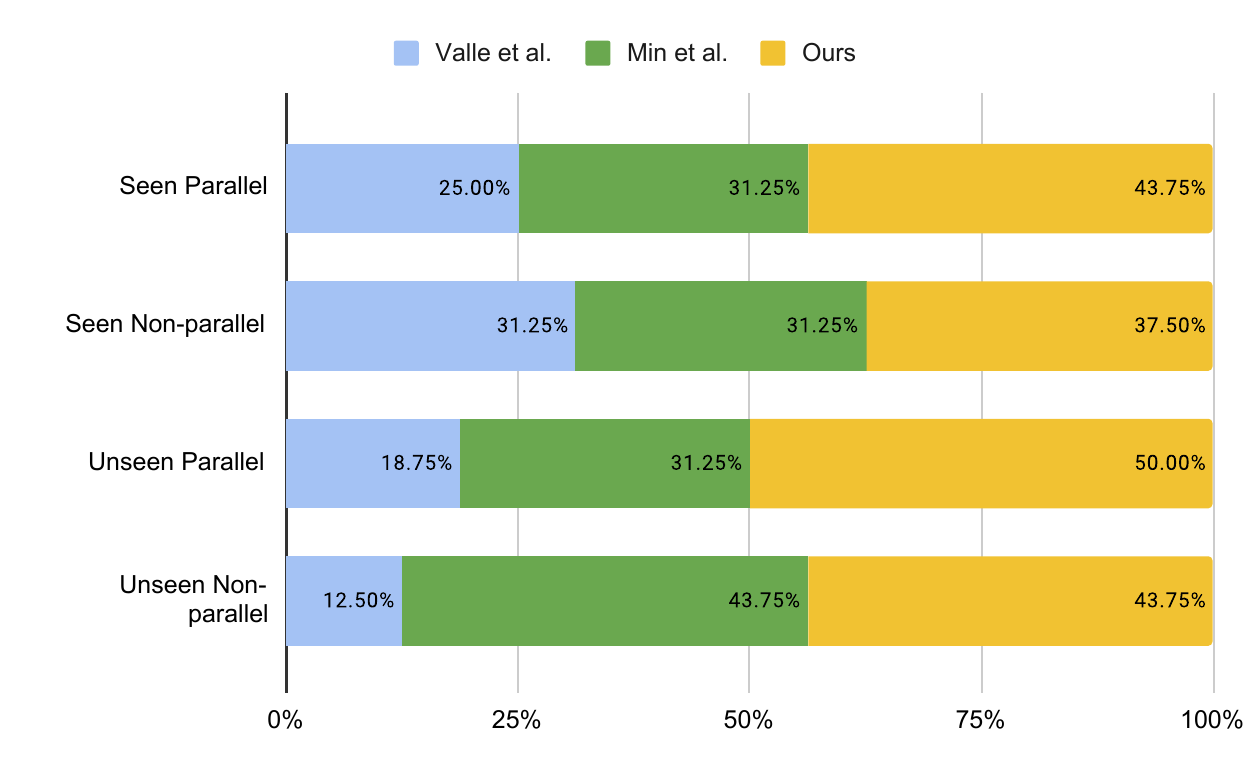}
    \caption{ABX comparisons between generated audio from different models. Best seen in color.}
    \label{fig:abx}
\end{figure}

\subsection{Speaker and Text Fusion}
Effective injection and fusion of the previously generated speaker embedding into the original TTS pipeline is crucial, as input text must be spoken in the target speaker's voice. Conventionally, the speaker embedding is concatenated with either the linguistic encoder outputs or the mel-spectrogram generator inputs of the TTS backbone. However, simple concatenation of target speaker information with target input text may not be enough to sufficiently fuse the two components together.

To address this limitation, we employ Style-Adaptive Layer Normalizations (SALN; \cite{SALN}) that involves two inputs: speaker embedding $s$, and hidden vector $i$ with dimensionality $H$ obtained from the linguistic encoder. First, $i$ is normalized using mean and standard deviation calculations, yielding normalized output $x$. Subsequently, a fully connected layer predicts the gain $g$ and bias $b$ with regards to the speaker embedding $s$ (Fig.~\ref{fig:main_arch}-(c)). This process is summarized as follows:

\begin{equation}
SALN(i, s) = g(s) \cdot x + b(s)
\end{equation}

Since the target text $x$ is adaptively adjusted in accordance with target speaker $s$, this facilitates a more effective fusion of the input text with the speaker's distinctive traits, ultimately leading to an improved quality of voice generation.

\subsection{Loss Function}

Our model is trained end-to-end in an unsupervised manner using a summation of different losses. Specifically, L1 distance is used to minimize the difference between the ground truth $M_n$ and synthesized mel-spectrogram $\hat{M}_n$ representations. In addition, we use L2 losses between the ground truth and predicted pitch ($P$), energy ($E$), and duration ($D$) information, which are introduced via the variance adapter. The full reconstruction loss is denoted as follows:

\begin{multline}
    \mathcal{L}_{recon} = \mathcal{L}_{mel}(M_n, \hat{M}_n) + \\ \mathcal{L}_{pitch}(P, \hat{P}) + \mathcal{L}_{energy}(E, \hat{E}) + \mathcal{L}_{dur}(D, \hat{D})
\end{multline}

\begin{figure}[t]
\centering
\includegraphics[width=0.71\linewidth,height=8.2cm]{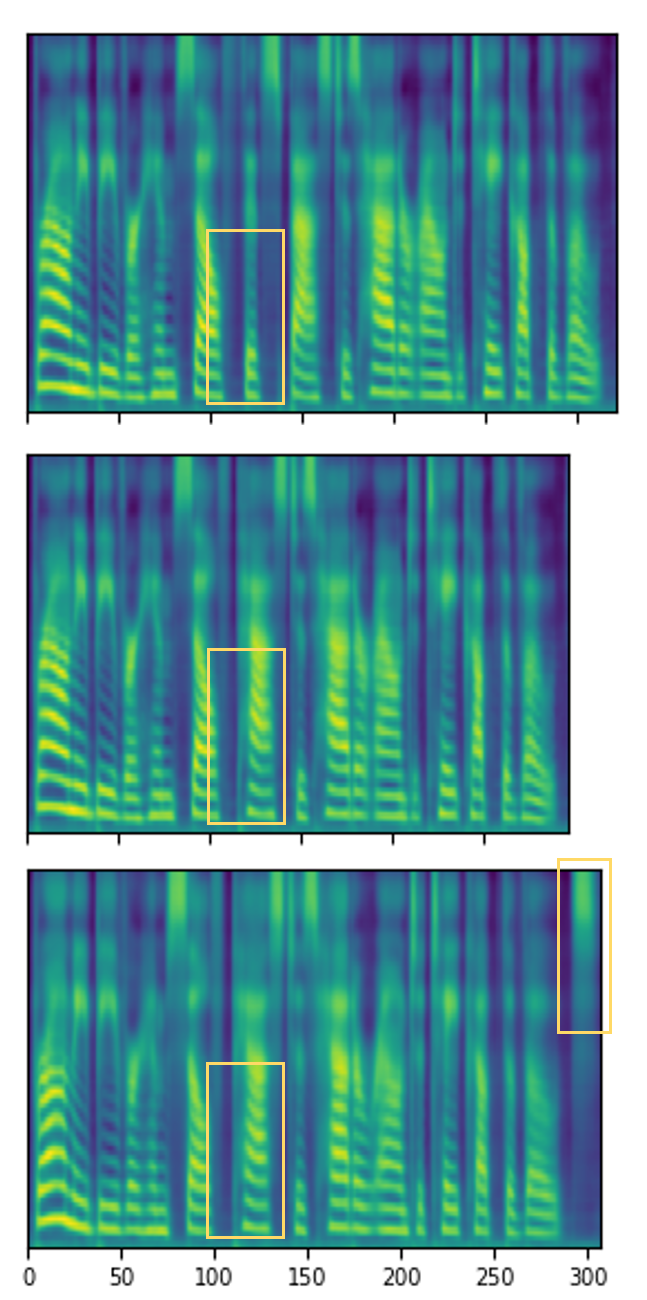}
\caption{Mel-spectrogram comparisons between \citeauthor{Mellotron}, \citeauthor{SALN}, and the proposed system (from top to bottom). Regions of notable disparities are highlighted in boxes. Notably, the uppermost sub-figure accentuates the instance of erroneously inserted enunciation. In addition, compared to the other models, voiceless alveolar fricative /s is correctly generated in the last sub-figure.}
\label{fig:melspec}
\end{figure}

\section{Experimental Settings}
\subsection{Dataset}
The benchmark LibriTTS train-clean-100 dataset \cite{LibriTTS} is used to conduct training and validation. This dataset is comprised of 53 hours of speech contributed by 247 people, with nearly equal representation from both genders (123 female and 124 male speakers). All audios are resampled at 22050 Hz and 16 bits, and mel-spectrograms are extracted with short-time Fourier transform \cite{STFT} using filter length of 1024, hop size of 256, window size of 1024, and 80 frequency bins. Prior to conducting experiments with this dataset, we applied a preprocessing step using the open-source Librosa library\footnote{https://librosa.org/doc/latest/index.html}, which involved trimming out any initial and final silences present in each audio sample. Additionally, we utilized the G2P library\footnote{https://github.com/Kyubyong/g2p} to convert the grapheme-based text into phonemes. 

\subsection{Evaluations}
We compare our model with two well-established baseline models, Mellotron \cite{Mellotron} and Meta-StyleSpeech \cite{SALN}. Mellotron is based on the Tacotron2 \cite{Taco2} model, which is augmented with GSTs, speaker embeddings, and pitch contours to perform multi-speaker voice synthesis. Meta-StyleSpeech adopts meta-learning in order to better adapt to unseen speakers, thereby simulating episodic training. In order to assess our model in terms with these two baseline models, we employ both subjective and objective metrics to assess our methodology.

\begin{table}[t]
\centering
\resizebox{0.93\columnwidth}{!}{\begin{tabular}{l|ccc}
\hline
&\textbf{MCD ($\downarrow$)} & \textbf{MOS ($\uparrow$)} & \textbf{WER ($\downarrow$)} \\\hline
Without Negation & 7.44 & 3.30 ± 0.23 & 16.68\% \\ \hline
\textbf{With Negation} & 7.08 & 3.85 ± 0.12 & 15.53\%  \\
\hline
\end{tabular}}
\caption{Experimental results to verify the efficacy of generating CIF speaker embeddings with negated learning. Seen and unseen speaker scores are averaged across all metrics.}
\label{Subtraction}
\end{table}

\begin{table}[t]
\centering
\resizebox{0.93\columnwidth}{!}{\begin{tabular}{l|ccc}
\hline
& \textbf{MCD ($\downarrow$)} & \textbf{MOS ($\uparrow$)} & \textbf{WER ($\downarrow$)} \\\hline
Wav2Vec Layer 1 & 7.51 & 3.43 ± 0.15 & 16.42\% \\
Wav2Vec Layer 7 & 7.80 & 3.12 ± 0.19 & 16.79\%  \\
Wav2Vec Layer 12  & 7.77 & 3.02 ± 0.21 & 16.65\% \\ \hline
\textbf{Instance Norm}  & 7.08 & 3.85 ± 0.12 & 15.56\% \\
\hline
\end{tabular}}
\caption{Ablation experiments and comparisons between various content extractor models.}
\label{w2v}
\end{table}

The Mean Opinion Score (MOS) is a subjective evaluation metric in which human participants rate the synthesized audio's similarity to the ground truth audio using a Likert scale ranging from 1 to 5. A score of 5 indicates the highest degree of speaker similarity between the synthesized audio and the ground truth audio. For MOS evaluation, we utilize the Amazon Mechanical Turk\footnote{https://www.mturk.com/} platform, where participants compare the speaker similarity of an anonymized audio that is synthesized from one of three models (i.e., two baseline models and our proposed model) to a specified ground truth audio. Each participant answered 53 comparison questions, with an average participating time of approximately 40 minutes. Final MOS scores are calculated with a 95\% confidence interval. In addition, as a complement to model-specific MOS validations, we further conducted intra-model A or B tests (ABX) in which the same pool of evaluators were presented with three audio files, each synthesized from a different model. They were then asked to choose the audio file that best matched the speaker's voice in comparison to the reference ground truth audio.

The objective metrics that are employed in our evaluations are Mel Cepstral Distortion (MCD) and Word Error Rate (WER). MCD quantifies the difference between the $k$th Mel Frequency Cepstral Coefficient (MFCC) vectors of the synthesized $\hat{S}$ and the ground truth audio signals $S$ for frame $T$ (Equation \ref{MCD}). Specifically, we use the dynamic time warping (DTW) version to take differences in audio durations into account. Smaller MCD scores indicates that the synthesized audio is more similar overall with the ground truth audio. We utilize the open source pymcd library\footnote{https://github.com/chenqi008/pymcd}. 

\begin{equation} 
    MCD = \frac{10 \sqrt{2}}{ln10} \frac{1}{T} \sum_{t} \sqrt{\sum_{k=1}^{K} (S_{t, k} - \hat{S}_{t, k})^2}
    \label{MCD}
\end{equation}

Word Error Rate (WER) is a widely utilized metric in Automatic Speech Recognition (ASR) to evaluate the accuracy of word pronunciation in synthesized audio. While not typically used in TTS research, we include this metric in our evaluations in order to examine the potential amount of content information that has leaked into speaker representations during the acquisition of speaker features from the reference audio. This is feasible since lower WER rates are indicative of improved pronunciation, thereby implying reduced content leakage within our acquired speaker representations. To compute WER, we leverage OpenAI's pretrained Whisper large-v2 model \cite{Whisper} to transcribe synthetic audios generated from different models, and subsequently employ the jiwer package\footnote{https://github.com/jitsi/jiwer} to precisely calculate the error rates.

\begin{table}[t]
\centering
\resizebox{0.96\columnwidth}{!}{\begin{tabular}{ll|ccc}
\hline
\textbf{Stream} & \textbf{Fusion} & \textbf{MCD ($\downarrow$)} & \textbf{MOS ($\uparrow$)} & \textbf{WER ($\downarrow$)} \\\hline
{Single} & N/A & 7.52 & 3.49 ± 0.18 & 15.65\% \\\hline
\multirow{2}{*}{Multi} & Concat & 7.31 & 3.51 ± 0.27 & 15.84\% \\
& Attn. Pool & 7.08 & 3.85 ± 0.12 & 15.53\%  \\ 
\hline
\end{tabular}}
\caption{Single and multi-stream comparisons with different hypotheses fusion methods.}
\label{tab:streamNum}
\end{table}

\subsection{Training Details}
Adam optimization is utilized with hyperparameters $\beta_1 = 0.9$, $\beta_2 = 0.98$, and $\epsilon = 10^{-9}$. To convert the generated mel-spectrograms into audio, we employ the HiFi-GAN \cite{HiFiGAN} vocoder. All experiments are conducted on a single RTX A6000 GPU with a batch size of 16, until step 300,000. On average, an experiment takes approximately 28 hours to complete. The total number of parameters of our proposed model is 30M. 

\section{Results and Discussions}
The results in Table 1 unequivocally demonstrate the superiority of our model, which surpasses the baseline systems across all three metrics for both seen and unseen speakers. Moreover, we provide visual evidence of the effectiveness of our negated feature learning process by conducting uniform manifold projections \cite{UMAP} of the final speaker embeddings (Fig.~\ref{fig:visualizations}). These visualizations clearly depict the enhanced clustering of utterances from the same speakers, validating our model's ability to better capture distinct speaker-specific representations, which ultimately aligns with higher speaker fidelity (MOS). 

While our model exhibits the highest MOS scores, evaluations were done separately for each model. Consequently, to achieve a more comprehensive assessment, we conduct inter-model comparisons through ABX across four possible conditions involving speaker visibility and parallelism of text utterances (Fig. \ref{fig:abx}). We find that while our proposed method generally has higher preference, clear model preferences manifest when both ground truth and synthesized audio share the same spoken context. Notably, the variance in preference between the two leading models diminishes from 12.50\% to 6.25\%, and from 18.75\% to 0\% for seen and unseen speaker conditions, respectively. In addition, we show in Figure \ref{fig:melspec} the acoustic fluctuations that occur across different models, which influences MCD and WER scores.

\begin{table}[t]
\centering
\resizebox{0.74\columnwidth}{!}{\begin{tabular}{l|cccc}
\hline
\textbf{Model} &  & \multicolumn{3}{l}{\textbf{MCD ($\downarrow$)}}  \\ \hline
\citeauthor{Mellotron}    & \multicolumn{4}{c}{6.60}                     \\ \hline
\citeauthor{SALN}         & \multicolumn{4}{c}{5.97}                     \\ \hline
\multirow{4}{*}{Proposed} & \multicolumn{1}{l|}{\diagbox[width=1.8cm, height=0.45cm]{}{}}  &\textbf{H2} & \textbf{H4} & \textbf{H8}\\ \cline{2-5}
                          & \multicolumn{1}{c|}{\textbf{D1}} & 5.59 & 5.71 & 5.64 \\ \cline{2-5} 
                          & \multicolumn{1}{c|}{\textbf{D2}} & 5.61 & 5.71 & 5.74 \\ \cline{2-5} 
                          & \multicolumn{1}{c|}{\textbf{D4}} & 5.69 & 5.77 & 5.64 \\ \hline 
\end{tabular}}
\caption{Multi-stream Transformer optimizations for seen speakers. The numbers next to head H and block depth D denote the precise parameters used.}
\label{seen-opt}
\end{table}

\begin{table}[t]
\centering
\resizebox{0.74\columnwidth}{!}{\begin{tabular}{l|cccc}
\hline
\textbf{Model} &  & \multicolumn{3}{l}{\textbf{MCD ($\downarrow$)}}  \\
\hline
\citeauthor{Mellotron}    & \multicolumn{4}{c}{8.97}                     \\ \hline
\citeauthor{SALN}         & \multicolumn{4}{c}{9.52}                     \\ \hline
\multirow{4}{*}{Proposed} & \multicolumn{1}{l|}{\diagbox[width=1.8cm, height=0.45cm]{}{}}  &\textbf{H2} & \textbf{H4} & \textbf{H8}\\ \cline{2-5}
                          & \multicolumn{1}{c|}{\textbf{D1}} & 8.57 & 8.85 & 8.70 \\ \cline{2-5} 
                          & \multicolumn{1}{c|}{\textbf{D2}} & 8.70 & 8.67 & 8.80 \\ \cline{2-5} 
                          & \multicolumn{1}{c|}{\textbf{D4}} & 8.84 & 8.82 & 8.95 \\ \hline 
\end{tabular}}
\caption{Multi-stream Transformer optimizations for unseen speakers.}
\label{unseen-opt}
\end{table}

\subsection{Impact of Negated Feature Generation}
Disentangled CIF speaker embeddings were achieved by negating content information from the full acoustic representation of the reference audio. To assess the impact of negation, we trained a separate model that bypassed this process and directly fed the input reference audio into the multi-stream Transformer architecture. However, as seen in Table \ref{Subtraction}, this yields lower voice similarity (MOS) and overall audio fidelity (MCD). Moreover, significantly increased WER infers increased content leakage in speaker representations, which leads to pronunciation instability due to conflicting contextual information from the reference audio and the actual target text. Hence, negation plays a crucial role in generating accurate CIF embeddings. 

While the effectiveness of negation has been established, the adequacy of instance normalization in extracting content information remains to be validated. To address this, we explore an alternative approach using a pretrained wav2vec 2.0 base model \cite{wav2vec2} as the context extractor. Specifically, embeddings outputted from the 1st, 7th, and 12th layers are used since each layer of wav2vec encapsulate different speech properties such as local acoustic features and phonetic content  \cite{wav2vec2_val}. Table \ref{w2v} reveals that using the first layer of wav2vec is most effective. Nevertheless, our method of using instance normalization proves more effective than the different variants of the pretrained wav2vec model.

\begin{figure}[t]
    \centering
    \includegraphics[width=0.9\linewidth, height=5cm]{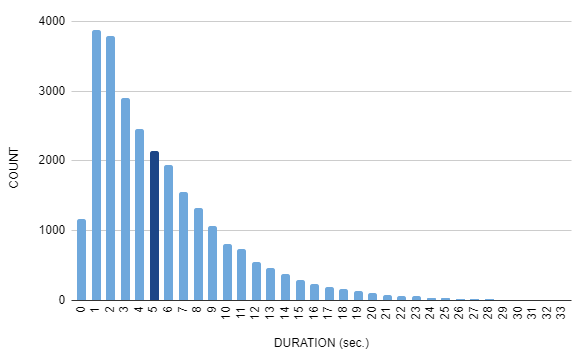}
    \caption{Long-tail distributions of the audio durations in the training dataset. The average duration for approximately 26,500 audio samples is 5.37 seconds (indicated in dark blue). Minimum and maximum audio lengths are 0 and 33 seconds, respectively.}
    \label{fig:audioStatistics}
\end{figure}

\subsection{Utilization of Multiple Hypotheses}
To assess the efficacy of multi-stream Transformers, we train three separate models. The first model incorporates a linear single-stream pathway to produce a single hypothesis. Recognizing the potential effect that multi-hypotheses fusion methods may have on performance, for fairer comparison, we developed a second model. This model employs multi-stream Transformers, but uses na\"ive concatenation to combine multiple hypotheses together to create the ultimate speaker embedding. A comparison between these two models demonstrates that even with na\"ive concatenation, multi-stream Transformers tends to surpass single-stream model performance (Table \ref{tab:streamNum}). Furthermore, to determine the most effective way to fuse multiple hypotheses, we compare the second model, which uses concatenation, with a third model (i.e., our proposed system) that applies attention pooling. Our findings reveal that attention pooling proves to be a more effective approach compared to simple concatenation.

We further conducted optimization experiments for the multi-stream Transformer component of our scheme by exploring various head settings, as well as the number of blocks used within each parallel stream (Tables ~\ref{seen-opt}, \ref{unseen-opt}). Interestingly, we discovered that the best performance was achieved with minimal head numbers and blocks. This observation could be attributed to the characteristics of our training dataset. As illustrated in Fig.~\ref{fig:audioStatistics}, a significant proportion (53.57\%) of the dataset have durations less than 5 seconds. Given the brevity of the audios and the fact that multi-stream Transformers conduct computations on a CIF speaker embedding that has already undergone content information pruning, it becomes more effective to adopt a smaller number of heads and blocks. Despite the increase in MCD scores when increasing the number of heads and/or blocks, it is important to note that all scores, both for seen and unseen speakers, outperform the baseline models.

\begin{table}[t]
\centering
\resizebox{0.92\columnwidth}{!}{\begin{tabular}{l|ccc}
\hline
& \textbf{MCD ($\downarrow$)} & \textbf{MOS ($\uparrow$)} & \textbf{WER ($\downarrow$)} \\\hline
Enc. Only & 8.66 & 3.00 ± 0.19 & 17.47\% \\ 
Dec. Only & 8.59 & 2.92 ± 0.19 & 17.56\%  \\ \hline
\textbf{Enc. \& Dec.} & 7.08 & 3.85 ± 0.12 & 15.53\%  \\\hline
\end{tabular}}
\caption{Speaker knowledge integration experiments. Optimal performance is observed when text and speaker are combined together at the fusion encoder, and speaker representations are additionally infused in the mel decoder.}
\label{SALN}
\end{table}

\subsection{Speaker Knowledge Infusion}
Strategic placement and frequency of fusing speaker knowledge within the basic TTS pipeline are crucial considerations. Thus, we additionally experiment with injecting speaker representations either exclusively into the TTS encoder or decoder. Our findings reveal that the most effective method is to incorporate speaker knowledge into both the encoder and decoder (Table \ref{SALN}). Also notable is that when speaker information is only integrated at the decoder, MCD slightly decreases, but MOS and WER scores are worse than encoder-exclusive fusion. This indicates that early fusion of text and speaker attributes is more important.

\section{Conclusion}
In this paper, we proposed for the first time, a novel negation feature learning scheme that could effectively extract speaker representations from a reference audio with less content leakage. Furthermore, we improve speaker fidelity through multi-stream Transformers by generating multiple hypotheses, which are all taken into account with attention pooling to generate the final speaker embedding. Through adaptive layer normalizations, target speaker attributes were fused into both the encoder and decoder of the baseline TTS model. Experiments have proven that our negated feature generation and learning methodology outperforms other well-known baseline models in terms of zero-shot speaker adaptation.

\section{Acknowledgments}
This research is supported by Culture Technology R\&D Program through the Korea Creative Content Agency funded by Ministry of Culture, Sports and Tourism (Development of contents meta-verse based on XR and AI, R2021040136) and by Institute of Information \& communications Technology Planning \& Evaluation (IITP) grant funded by the Korea government(MSIT) (No.2019-0-01906, Artificial Intelligence Graduate School Program(POSTECH)).

\bibliography{aaai24}

\end{document}